\documentclass[10pt,conference]{IEEEtran}
\IEEEoverridecommandlockouts

\usepackage[table]{xcolor}
\usepackage{url} 
\usepackage{framed}
\usepackage{mdframed}
\usepackage{listings}
\usepackage{multicol}
\usepackage{amssymb}
\usepackage{lipsum}
\usepackage{adjustbox}
\usepackage[font=bf,skip=\baselineskip]{caption}
\usepackage{tabularx}
\usepackage{seqsplit}
\usepackage{multirow}
\usepackage{booktabs}
\usepackage{graphicx}
\usepackage{hyperref}
\usepackage{subcaption}
\usepackage{makecell}
\usepackage[table]{xcolor}
\usepackage[ruled,vlined,linesnumbered]{algorithm2e}
\usepackage{amsmath}
\usepackage{tikz}
\usepackage{tcolorbox}
\usepackage{threeparttable}
\usepackage{bbm}

\usepackage{breakurl} 
\usepackage{boldline}
\usepackage[ruled,vlined]{algorithm2e}

\usepackage{pifont}
\usepackage{balance}

\usepackage{enumitem}
\usepackage[htt]{hyphenat}
\usepackage{array}

\definecolor{customblue}{HTML}{006ca6}
\definecolor{customgreen}{HTML}{009264}
\definecolor{custombrown}{HTML}{ff3d00}
\AtEndPreamble{
\usepackage{hyperref}

}

\newcommand{\tool}{\textsc{ArtifactCopilot}}

\newcommand{\rubric}{\textsc{ArtifactGuide}}

\begin{document}

\title{Agent-Based Software Artifact Evaluation}

\author{
\IEEEauthorblockN{
Zhaonan Wu\textsuperscript{1},
Yanjie Zhao\textsuperscript{1}\ding{41},
Zhenpeng Chen\textsuperscript{2},
Zheng Wang\textsuperscript{3},
and Haoyu Wang\textsuperscript{1}
}
\IEEEauthorblockA{\textsuperscript{1}Huazhong University of Science and Technology, Wuhan, China\\
\{zhaonanwu, yanjie\_zhao, haoyuwang\}@hust.edu.cn}
\IEEEauthorblockA{\textsuperscript{2}Tsinghua University, Beijing China, zpchen@tsinghua.edu.cn}
\IEEEauthorblockA{\textsuperscript{3}Independent Researcher, hackerzheng666@gmail.com}
\thanks{\ding{41} Corresponding author: Yanjie Zhao (yanjie\_zhao@hust.edu.cn).}
}

\maketitle

\begin{abstract}
Artifact Evaluation (AE) has become a standard mechanism for linking software engineering papers to their supporting artifacts, but growing submission volume makes purely manual AE increasingly difficult to sustain. Although recent advances in LLM-based agents make AE automation increasingly plausible, current descriptive badge policies define badge semantics but provide no executable verification criteria,  leaving both human reviewers and agents without a detailed basis for judgment. To address this gap, we construct \rubric{}, a structured scoring rubric grounded in ACM policy, expert-informed calibration, and artifact-based validation, and we design \tool{}, an agent collecting review evidence following a fixed sequence under \rubric{} and deriving the final badge decision from accumulated evidence.

We evaluate our framework on 60 real artifacts from recent software engineering conferences using human-adjudicated badges as reference. The results show that \rubric{} improves the AE performance of coding agents over official ACM badge-policy prompts, increasing three-run mean exact badge agreement by 10.55 to 28.34 percentage points. Across all evaluated systems and prompting protocols, \tool{} achieves the highest badge-level agreement at 70.56\% and is the only system that completed all repeated runs successfully while producing a review report in every run. A controlled user study with 8 experienced researchers suggests that \tool{} reports improve reviewer confidence, help reviewers locate evidence, and understand evaluation scope more clearly. Further analysis translates insights from automated AE into practical guidance for designing higher-quality artifacts with clearer review routes, more explicit claim-to-output links, and more concrete reuse paths.
\end{abstract}


\section{Introduction}
\label{sec:introduction}

Since artifact evaluation (AE) was first deployed at ESEC/FSE in 2011~\cite{krishnamurthi2013artifactEvaluation}, it has been widely adopted across software engineering, programming languages, and broader computer science venues~\cite{winter2022retrospectiveAE}. Nevertheless, neither its criteria nor its processes have kept pace with this proliferation. As submission and artifact volumes continue to grow, artifact review remains heavily reliant on manual effort, rendering the process increasingly unsustainable at scale~\cite{treude2026rethinking, artisan2026, sigmobile_artifact_guidelines, eurosys2025lessonsLearned}. Recent advances in LLM-based software agents have substantially expanded the scope of automated software engineering tasks. Software agents can navigate repositories, execute commands, configure environments, and record execution traces~\cite{yao2023react, yang2024sweagent, wang2025openhands}, offering a promising path toward automated AE. However, prior work has demonstrated that when LLMs or agents are deployed for evaluation tasks, performance depends critically on whether the rubric is sufficiently explicit and operational~\cite{prather2025rubric}.  Therefore, automating AE with agents first requires a detailed rubric. In a similar vein, empirical findings in AE corroborate this requirement for a more detailed rubric. Hermann et al.~\cite{hermann2020communityExpectations} surveyed artifact evaluation committee members across major conferences and found that existing official policies fail to yield clear and consistent criteria for acceptable artifacts, while community expectations remain systematically unreflected in calls for artifacts, leaving reviewers without adequate guidance for badge decisions.

These observations point to a more fundamental issue: the AE community still lacks a concrete and operational rubric. As AE faces growing scalability challenges and agents are increasingly considered as a possible solution~\cite{he2026llmAE, artisan2026}, the need for such a rubric becomes even more pressing. We therefore argue that a concrete and operational rubric is an essential missing component of AE, for both human reviewers and agents.

Acting on this thesis, we first design \rubric{}, an operational artifact-evaluation rubric that bridges the gap between descriptive policy and badge decisions. \rubric{} translates badge semantics into an executable review standard through an availability gate, five assessment dimensions, and ordered intervention and evidence levels. We construct this rubric through three phases, deriving an initial version from official ACM policy and published AE guidelines, calibrating judgment boundaries through semi-structured interviews with 20 experienced artifact evaluators, and validating the revised rubric through manual reviews of real artifacts. On top of \rubric{}, we then design \tool{}, a specialized agent for rubric-guided artifact review that takes the paper and its corresponding artifact as input, uses phase-specific skills~\cite{anthropic2026agentskills} to guide each review stage, produces structured evidence records, and derives final badge recommendations from them.

We then compared \tool{} against three coding agents and a project-execution agent under three review protocols with increasing structure, including official ACM badge policy~\cite{acmBadgingV11}, \rubric{}, and Skills+\rubric{} review. We also analyzed repeated-run stability and conducted a controlled user study with eight experienced researchers. The results show that \rubric{} improves the AE performance of coding agents over official ACM badge-policy prompts, increasing three-run mean exact badge agreement by 10.55 to 28.34 percentage points for OpenHands~\cite{wang2025openhands}, Cline~\cite{cline2024}, and Codex~\cite{openai2026codex}. Across all evaluated systems and protocols, \tool{} achieves the highest three-run mean exact badge agreement at 70.56\%. It is also the only evaluated system that completed all repeated runs successfully and produced a review report in every run.  The user study further suggests that \tool{} reports help reviewers locate evidence more effectively and recognize evaluation limitations more clearly.

Our further analysis shows that automated AE is useful not only for evaluating artifacts, but also for clarifying how higher-quality artifacts should be designed. Artifacts that can be reviewed stably by an agent tend to expose clearer review routes, more explicit claim-to-output links, and more concrete bounded reuse paths, whereas agent uncertainty often reflects underspecified artifact design.

Our main contributions are as follows.

\begin{itemize}
\item We design \rubric{}, a concrete and operational AE rubric that fills the gap between descriptive badging policy and actual badge decisions.
\item We design \tool{}, a specialized agent for rubric-guided AE review, and show through systematic evaluation on 60 real artifacts that it achieves the highest badge-level agreement among all compared systems, reaching 70.56\% exact badge agreement. 
\item We show that \rubric{} can be reused by coding agents as an effective external review structure, consistently improving their AE performance over official policy alone.
\item We provide an in-depth analysis of failure modes in automated AE and derive practical implications for both reviewer-oriented AE automation and artifact design.
\end{itemize}

\section{Background and Motivation}
\label{sec:background}

\subsection{Artifact Evaluation}
\label{subsec:ae}
The ACM artifact badging policy defines \emph{Available}, \emph{Functional}, and \emph{Reusable} badges as a shared institutional framework for artifact quality~\cite{acmBadgingV11}. Since its pilot at ESEC/FSE 2011~\cite{krishnamurthi2013artifactEvaluation}, AE has become standard practice across major software engineering venues~\cite{icse2026ae, ase2026ae, fse2026ae, issta2026ae}. Yet the institutionalization of AE has exposed a  tension between badging policy and review practice. Existing criteria remain descriptive rather than operational, specifying what each badge represents without defining how reviewers should verify it. This gap leaves individual reviewers to reconstruct judgment standards from personal experience, which undermines consistency and complicates scaling. Even when artifacts execute successfully, reviewers face unsupported decisions about whether observed deviations still permit reasonable claim support~\cite{eurosys2025lessonsLearned}. The resulting subjectivity not only threatens evaluation reliability but also drives resource demands. Each Artifact Evaluation Committee (AEC) member is typically assigned only 1--3 artifacts~\cite{eurosys2025lessonsLearned}, and EuroSys 2025 required 98 committee members to sustain this practice. As submission volume grows, this labor-intensive model is becoming unsustainable~\cite{treude2026rethinking, artisan2026}.

\subsection{LLM-based Agents}
\label{subsec:agent}

The continuing improvement of LLM capabilities has driven the formation of a new automation paradigm, namely agent systems centered on LLMs~\cite{Wang2023ASO, xi2025rise, xu2025llm}. LLM-based agents extend single-shot inference by accumulating context across steps, invoking external tools, and incorporating execution feedback into the reasoning process~\cite{Wang2023ASO, yao2023react}, which has been applied in SE to code repair, vulnerability detection, and test generation~\cite{bouzenia2025repairagent, OGHarn2025, mundler2024code}, consistently demonstrating advantages on tasks requiring cross-step reasoning and environmental feedback~\cite{yang2024sweagent, wang2025openhands}. Critically, agent execution trajectories constitute traceable records of task states~\cite{ICLR2024_7274ed90}, a property that recent work has begun to exploit for evaluation itself, replacing static assessment with executable dynamic processes~\cite{zhuge2024agentAsJudge, gou2025mind2web2}. This traceability directly addresses AE's core need for evidence-based judgment.

\begin{figure*}[t]
\centering
\includegraphics[width=\linewidth]{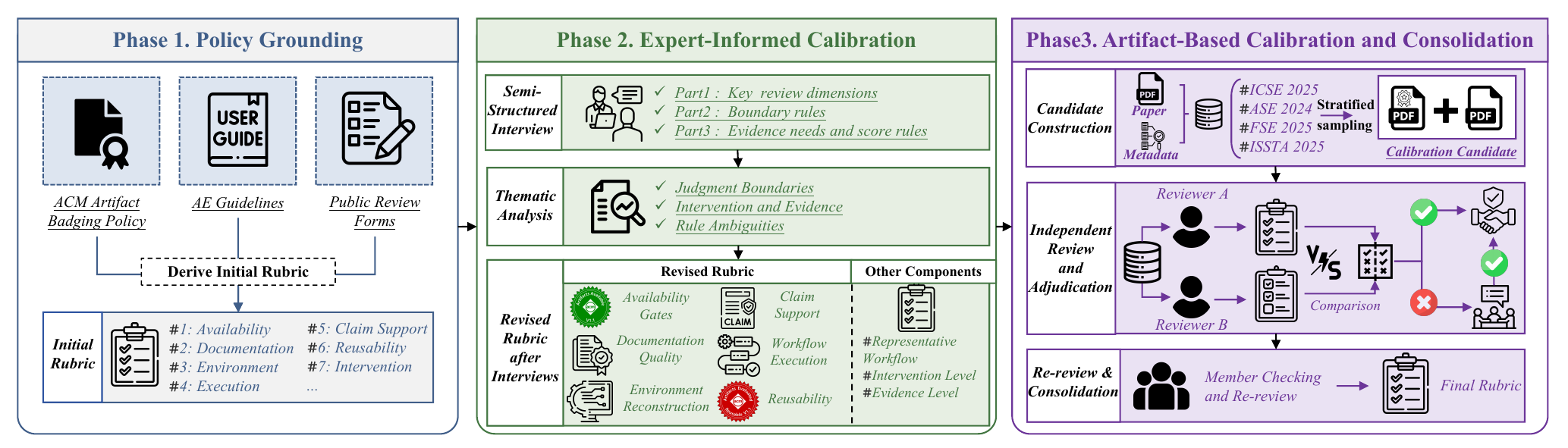}
\caption{The Workflow of the Construction of \rubric{}.}
\label{fig:rubric}
\end{figure*}

\subsection{Problem Formulation}
\label{subsec:problem-formulation}
These observations suggest that AE automation requires both explicit standards and executable enforcement. We define automated AE as an agent-supported, rubric-based evidence review task. The rubric provides the what, decomposing badge standards into explicit inspection items, evidence requirements, and decision thresholds. The agent provides the how, executing these items as bounded environment interactions and accumulating structured evidence. Neither alone suffices. A rubric without execution remains static inference. An agent without rubric guidance operates without evaluation criteria. Together, they shift AE from subjective badge assignment to traceable, auditable evidence review.

\section{The Design of \rubric{}}
\label{sec:rubric}

\subsection{Scope and Design Principles}
\label{subsec:rubric-scope}
We design \rubric{} as a multi-level review framework that operationalizes descriptive badge standards. By specifying the concrete items to be checked, the evidence to be recorded, and the logic by which evidence supports conclusions, it provides a unified basis for artifact evaluation without redefining ACM badges or superseding venue-specific policies. \rubric{} is designed around three interlocking principles. First, \textbf{policy alignment} requires that the scope and criteria of \rubric{} remain faithful to the official definitions of the badges. Second, \textbf{analytic decomposition} requires that \rubric{} decomposes badge judgment into separately checkable intermediate dimensions. This design draws on the established distinction in scoring rubric design between analytic and holistic rubrics~\cite{jonsson2007use}. Third, \textbf{evidence grounding} requires that each dimensional judgment be tied to concrete objective evidence, making each determination traceable.

\subsection{Construction of \rubric}
\label{subsec:rubric-derivation}

Based on the established methodology for rubric construction proposed by~\cite{asee_peer_41236, rudolph2025don}, \rubric{} was constructed through a three-phase process. We first conducted policy grounding to establish the rubric's policy basis and then revised the draft through expert calibration. Finally, we consolidated the rubric through manual review of real public artifacts. \autoref{fig:rubric} illustrates this construction process.

\textbf{Phase 1. Policy Grounding.}
We built the initial version of \rubric{} based on the ACM artifact badging policy~\cite{acmBadgingV11}, combined with specific requirements from recent AE guidelines of ICSE, ASE, FSE, and ISSTA. We further consulted publicly available review forms and guidelines to capture signals that are commonly omitted in formal badge descriptions but regularly attended to in actual review practice, such as whether the artifact is obtainable and has a clear entry point, and whether installation and execution instructions are complete. Synthesizing these sources, the initial version established the basic framework for subsequent iterations, covering availability, documentation, environment, execution, claim support, reusability, and reviewers' interventions.

\textbf{Phase 2. Expert-Informed Calibration.}
While the initial version captured review objects that commonly arise in AE, it fell short of reflecting what real reviewers actually prioritize in practice.  We therefore revised it through expert-informed calibration, interviewing twenty participants with artifact-related experience. The participant pool covered three forms of experience relevant to AE: prior service on artifact evaluation committees, paper submission and artifact preparation, and substantial hands-on experience in reproducing published research artifacts. These categories were not mutually exclusive, and several participants had experience across multiple roles. This composition allowed the calibration to reflect reviewer-side judgment, author-side preparation, and practical reproduction work. The final five interviews yielded no new dimensions, suggesting that the core themes had stabilized, so we concluded recruitment. 

Interviews were conducted as one-on-one semi-structured sessions. Participants were first asked to describe key review dimensions based on their own experience. Then the initial version was presented, and boundary rules were discussed, such as the acceptable scope of reviewer interventions and the distinction between execution success and result support. After the interviews, we conducted thematic analysis~\cite{braunclarke2006thematic} of the interview records. Two authors independently completed initial coding and resolved coding differences through discussion. A feedback item was treated as a generalizable rule only when it recurred across multiple participants' descriptions, rather than being tied to a particular artifact or an individual participant's preference. Compared with the initial version, the revised \rubric{} established clearer judgment boundaries across the Availability gate and the five dimensions of documentation quality, environment reconstruction, workflow execution, claim support, and reusability. Furthermore, we summarized the concepts of representative workflow, evidence levels, and intervention levels.
%As the interviews progressed, expert feedback converged on refining these boundaries without triggering substantial revisions to the core dimension structure, indicating that the overall structure of \rubric{} had stabilized.

\textbf{Phase 3. Artifact-Based Calibration and Consolidation.}
Phase 2 calibrated key judgment dimensions and evidence requirements through interviews, but the feedback remained limited to participants' discussions of prior experience and hypothetical scenarios. It was insufficient to demonstrate that the revised rubric could be applied consistently in real artifacts. We therefore further validated the revised rubric through manual review of real public artifacts. We constructed a calibration set from 642 research track papers across ICSE 2025, ASE 2024, FSE 2025, and ISSTA 2025, selecting those with publicly accessible artifacts and stratifying by official badge status. We used ASE 2024 rather than ASE 2025 because ASE 2025 did not include an artifact evaluation track. 

\newcommand{\lvl}[1]{\textbf{#1}}
\newcommand{\mylabel}[1]{\textbf{#1:}~}
\begin{table*}[htbp]
\centering
\caption{Gate conditions and dimension-level scoring criteria of \rubric{}. 
Intervention levels (I0–I4) and evidence levels (E0–E3) are defined in \autoref{tab:rubric-tiers-badges}.}
\label{tab:rubric-gate-dimensions}
\scriptsize
\renewcommand{\arraystretch}{1.0}
\setlength{\tabcolsep}{2.5pt}
\begin{tabularx}{\linewidth}{@{}>{\bfseries}p{0.14\textwidth} p{0.84\textwidth}@{}}
\toprule
Component & \textbf{Scoring Criteria} \\
\midrule
Availability gate &
\mylabel{GC1} publicly accessible and downloadable; 
\mylabel{GC2} README or other entry exists; 
\mylabel{GC3} related to the paper and contains material beyond the text. \\
\midrule
Documentation\hspace{6pt}Quality &
\mylabel{R} no usable entry, no environment or workflow instructions; 
\mylabel{WR} entry exists, but key information is largely missing; 
\mylabel{WA} goal, setup, workflow, I/O, and paper mapping documented with minor inference gaps; 
\mylabel{A} core guidance plus structure, dependencies, resources, and failure boundaries. \\
\midrule
Environment\hspace{0pt}Reconstruction &
\mylabel{R} no usable path, failed setup, or depends on I4-level intervention; 
\mylabel{WR} setup described but requires substantial repair; 
\mylabel{WA} reaches evaluable state with I0--I2 or semantic-preserving I3; 
\mylabel{A} stable, isolated, repeatable setup with clear dependency information. \\
\midrule
Workflow\hspace{40pt}Executability  &
\mylabel{R} no workflow or no checkable execution state; 
\mylabel{WR} candidate fails, lacks valid output, or only runs trivial commands; 
\mylabel{WA} produces E2/E3 valid output without I4 intervention; 
\mylabel{A} completes stably with I0 or I1 and checkable output. \\
\midrule
Claim support &
\mylabel{R} no E2/E3 evidence or no explainable paper mapping; 
\mylabel{WR} output maps weakly or peripherally to paper; 
\mylabel{WA} maps to at least one method, claim, result, table, figure, or experiment; 
\mylabel{A} supports the main result or full reproduction with a traceable path. \\
\midrule
Reusability &
\mylabel{R} no functional base, or no reuse entry identifiable; 
\mylabel{WR} reuse entry exists, but transfer information is insufficient; 
\mylabel{WA} transfer validation succeeds on new input, configuration, or scenario; 
\mylabel{A} reuse path clear, low-intervention, repeatable. \\
\bottomrule
\end{tabularx}
\end{table*}

\begin{table*}[htbp]
\centering
\caption{Representative-workflow checklist, intervention and evidence levels, and badge-derivation rules of \rubric.}
\label{tab:rubric-tiers-badges}
\scriptsize
\renewcommand{\arraystretch}{1.0}
\setlength{\tabcolsep}{2.5pt}
\begin{tabularx}{\linewidth}{@{}>{\bfseries}p{0.15\textwidth} X@{}}
\toprule
Component & \textbf{Criteria} \\
\midrule
Representative\hspace{10pt}workflow &
\mylabel{RW1} grounded in README, docs, paper appendix, author script, or official workflow;
\mylabel{RW2} invokes an evaluation-relevant program, script, notebook, service, demo, benchmark, analysis, test, or replication path;
\mylabel{RW3} produces interpretable valid output;
\mylabel{RW4} maps output to a paper method, result, claim, table, figure, or artifact target;
\mylabel{RW5} states coverage limit if not full reproduction. \\
\midrule
Intervention level &
\mylabel{I0} run as documented;
\mylabel{I1} install dependencies, set environment variables, or choose compatible versions;
\mylabel{I2} fix paths, commands, startup order, or wait time;
\mylabel{I3} semantic-preserving configuration, wrapper, non-core script, or compatibility change;
\mylabel{I4} semantic change to algorithm, model, metric, experiment, result generation, or core data processing. \\
\midrule
Evidence level &
\mylabel{E0} no executable evidence;
\mylabel{E1} static or precomputed materials only;
\mylabel{E2} reviewer runs representative workflow and obtains valid output;
\mylabel{E3} reviewer runs full or near-full main workflow and obtains paper-facing result objects. \\
\midrule
Badge rules &
\mylabel{Available} GC1--GC3 pass;
\mylabel{Functional} Available, D2--D4 $\ge$ WA, E2 or above, no I4;
\mylabel{Reusable} Functional, both D1 and D5 $\ge$ WA, transfer validation succeeds, no I4. \\
\bottomrule
\end{tabularx}
\end{table*}

We sampled 16 badged and 8 unbadged artifacts, a 2:1 ratio intended to test the rubric against both categories. Two authors independently reviewed the 24 artifacts using the revised rubric, achieving 83.3\% agreement on the availability gate, 74.2\% on 120 dimension-level judgments, and 54.2\% on final badge types. The badge-level disagreements exposed four boundary ambiguities, including whether the availability gate checks completeness or mere accessibility, whether Functional requires full reproduction or accepts representative execution, whether Reusable demands independent input construction or permits provided demo scripts, and whether external dependencies block scoring or are recorded as limitations. A third author adjudicated these cases to resolve scores and identify root causes. After all calibration artifacts were fully evaluated and adjudicated, \rubric{} was finalized. Revisions clarified the evidence-to-claim mapping, delineated the boundary between workflow execution and result support, and made reviewer action recording explicit to prevent conflating original artifact quality with post-intervention success. The two reviewers then re-reviewed all 24 calibration artifacts, eliminating residual disagreements caused by rule ambiguity and reaching full agreement. The finalized version was submitted to the interview participants, and none raised objections to the final design of \rubric{}.

\subsection{Structure of \rubric}
\label{subsec:rubric-structure}

The finalized \rubric{} first applies an Availability gate to determine whether an artifact satisfies the minimum material conditions for substantive review. It then evaluates five core dimensions, D1 through D5, covering documentation, environment reconstruction, workflow execution, claim support, and reusability. Each dimension follows a four-level scale of reject (R), weak reject (WR), weak accept (WA), and accept (A), adopted from traditional academic peer review~\cite{vijaykumar2018common}, with WA serving as the minimum positive threshold for satisfying the corresponding badge condition. The five dimensions are grouped by badge tier. D2 (environment), D3 (execution), and D4 (claim support) determine whether an artifact reaches the Functional threshold, while D1 (documentation) and D5 (reusability) are additional requirements for the Reusable badge. This grouping reflects the empirical finding that an artifact must first be executable before its reusability can be assessed, and that documentation quality is not necessary for functional correctness.
To prevent any runnable command from being treated as valid review evidence, \rubric{} also defines representative workflow conditions (RW1 to RW5), evidence levels (E0 to E3), and intervention levels (I0 to I4). The intervention levels distinguish original artifact quality from reviewer-assisted success, ensuring that post-intervention execution does not inflate the badge recommendation. The evidence levels ensure that only reviewer-executed outputs (E2 or above) count toward badge decisions, excluding static materials or precomputed results. \autoref{tab:rubric-gate-dimensions} summarizes the availability gate and dimension scoring criteria, while \autoref{tab:rubric-tiers-badges} presents the supporting criteria for determining evidence validity and final badge assignment. Due to space constraints, the complete \rubric{} with illustrative examples is available in our open-source artifact~\cite{ArtifactCopilot2026}.

\section{\tool{}}
\label{sec:system}
\begin{figure*}[t]
\centering
\includegraphics[width=\linewidth]{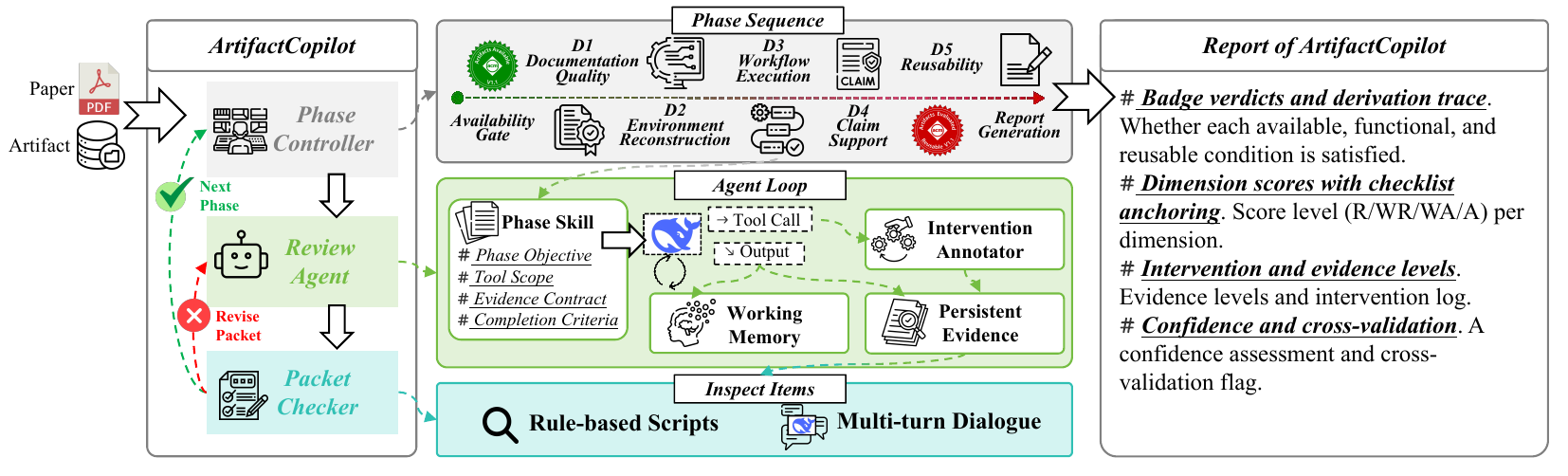}
\caption{Overview of \tool{}.}
\label{fig:agent}
\end{figure*}

\subsection{System Overview}
\label{subsec:system-overview}
The \tool{} module models AE as a phased evidence review process. \tool{} first checks artifacts in a fixed phase sequence, generating verifiable evidence packages at each phase. The system only proceeds to the next phase after the phase evidence passes the check. The final badge is derived from persistently accumulated evidence rather than from the agent's immediate response. As shown in \autoref{fig:agent}, the input to \tool{} is the paper and its corresponding artifacts. The \textbf{Phase Controller} and \textbf{Packet Checker} are responsible for the phase progression and acceptance. The execution of each phase is handled by the \textbf{Review Agent}, which invokes tools under the constraints of the current phase skill, observes the output, and records the execution evidence and intervention actions during the process. Upon completion of the final phase, \tool{} outputs a structured review report.

\subsection{Phase Orchestration}
\label{subsec:phase-runtime}

The control layer of the \tool{} module is responsible for transforming \rubric{} from static scoring criteria into an executable review process. The Phase Controller enforces a fixed phase sequence, starting with the Available Gate. Within each phase, the Review Agent can choose specific tool calls and inspection paths based on the specific characteristics of the current artifact; however, the phase objectives, permitted operations, evidence required for phase completion, and completion conditions are all defined by the phase skill. Therefore, \tool{} can adapt to different artifacts while avoiding the invention of review methods in each evaluation.

The Packet Checker verifies the evidence packet before phase transition. At the end of each phase, the Review Agent must submit a packet containing the required evidence artifacts and phase conclusions. The checker applies two lightweight checks. First, it runs rule-based validation to confirm that the required files exist and that the packet is structurally complete. Second, it invokes a bounded dialogue verifier with a maximum of five turns to examine whether the stated phase conclusions are actually supported by the submitted logs and related documents. This second step is intentionally lightweight whose role is only to check evidence-conclusion consistency, not to re-review the artifact or produce new judgments. Only packets that pass both checks are committed and allowed to advance to the next phase. Otherwise, \tool{} returns to the current phase to supplement or revise the evidence.

It is important to emphasize that the Availability gate incorporates a unique early stopping branch. If an artifact fails the checks of GC1--GC3, the Phase Controller will automatically skip directly to the phase of report generation, avoiding excessive resource consumption on artifacts that do not meet the minimum evaluation criteria.

\subsection{Review Agent}
\label{subsec:review-agent}

\textbf{Skill-scoped execution.} The Review Agent is responsible for evidence collection and integration into report generation. Each phase is governed by a phase skill, a pre-written instruction artifact that specifies four elements: Phase Objective, Tool Scope, Evidence Contract, and Completion Criteria. Upon entering a new phase, the agent reads the phase skill and invokes tools within its Tool Scope to produce evidence that satisfies the Evidence Contract and Completion Criteria. These skills limit the agent's operational space. For example, the document quality phase primarily allows reading and observing the document structure, while the environment reconstruction phase allows dependency probing and environment configuration but strictly prohibits the execution of representative workflows. This progressive scoping prevents evidence contamination between phases. 

\textbf{Dual memory and persistent evidence.} The Review Agent maintains two types of memory to prevent temporary planning and partial observations from being conflated with auditable evidence. The first is working memory, which stores the current phase's transient state, including current tasks, recent observations, and pending execution plans. The second is persistent memory, which stores the phase evidence that is eligible for submission, including phase summaries, instruction logs, and intervention records. This information is permanently saved upon phase completion and serves as the direct source of the final report.

\textbf{Intervention annotation.} The Review Agent records actions that introduce reviewer-side assistance beyond the artifact's original instructions, such as installing missing dependencies, modifying paths or startup order, supplying external inputs, or applying compatibility-oriented fixes. Each such action is checked by a script-based intervention rule and logged with its purpose, intervention type, and anticipated downstream impact. The resulting record becomes part of persistent evidence, preventing the agent from retrospectively downplaying the impact of an intervention after observing successful execution.

\subsection{Structured Review Report}
\label{subsec:scoring-report}
The final phase produces a structured review report in both Markdown and JSON formats, which are structured to support human audit, deriving all claims from persistent evidence.

\textbf{Badge verdicts and derivation trace.} The first component details the badge verdicts and their derivation traces. For each Available, Functional, and Reusable badge, the report explicitly states the deterministic rule applied and links directly to the specific evidence artifacts that satisfy the gate conditions. For instance, a Functional badge decision cites the workflow execution log from dimension D3 and the claim mapping file from dimension D4. All references include hyperlinks to the corresponding run directories for immediate verification.
 
\textbf{Dimension scores with checklist anchoring.} The second component presents the dimension scores with checklist anchoring. For every score from dimension D1 to D5, the report enumerates the specific checklist items from \rubric{} that were satisfied or violated. Each item is directly linked to the corresponding command output or log excerpt. This granular linkage empowers reviewers to inspect the foundational evidence and substitute their own judgments where necessary.
 
\textbf{Intervention and evidence levels.} The third component is a consolidated ledger of intervention and evidence levels. It records all interventions from level I0 to I4 alongside their rubric justifications and assigns the appropriate evidence level from E0 to E3 for each dimension. Such transparency is crucial for badge decisions because the rubric disqualifies level I4 interventions from receiving Functional or Reusable badges.
 
\textbf{Confidence and cross-validation.} The fourth component provides reviewer-facing confidence cues and consistency flags. These signals do not directly affect the final badge decision, which is still derived from the deterministic rules of \rubric{}. Instead, they help reviewers identify cases that may require closer inspection. Confidence is estimated from the completeness of supporting evidence and the intervention burden recorded in the run. Cross-validation applies a small set of rule-based consistency checks, for example whether D3 $\ge$ WA is accompanied by E2 or above, whether D4 $\ge$ WA is supported by a successful execution result, and whether a Reusable recommendation is compatible with the Functional gate and D5 judgment. The report lists any resulting flags according to their severity.

\section{Evaluation}
\label{sec:evaluation}

We organize our evaluation around four research questions.

\begin{itemize}[leftmargin=*, nosep]
  \item \textbf{RQ1.} How closely does \tool{} align with human adjudication?
  \item \textbf{RQ2.} How does \tool{} compare with baseline agents under different review guidance?
  \item \textbf{RQ3.} How stable is \tool{}'s behavior across repeated runs?
  \item \textbf{RQ4.} How useful are \tool{}'s reports in AE review?
\end{itemize}

\subsection{Experimental Setup}
\label{subsec:eval-setup}
\begin{table*}[t]
  \centering
  \caption{RQ1: Alignment with human adjudication across badge, gate, and dimension levels.}
  \label{tab:rq1_summary}
  \small
  \begin{tabularx}{\textwidth}{@{}llXrrrrr@{}}
  \toprule
  \textbf{Layer} & \textbf{Item} & \textbf{Description} & \textbf{Acc.} & \textbf{Prec.} & \textbf{Rec.} & \textbf{F1} \\
  \midrule
  \textbf{Badge}
    & Exact match & Four-level badge
    & 70.56\% & 74.05\% & 72.05\% & 70.64\% \\
  \midrule
  \textbf{Gate}
    & Available  & GC1--GC3 pass
    & -- & 98.14\% & 97.53\% & 97.82\% \\
    & Functional & $\ge$Available + D2--D4$\ge$WA
    & -- & \textbf{98.55\%} & 70.59\% & 82.23\% \\
    & Reusable   & $\ge$Functional + D1,D5$\ge$WA
    & -- & 87.44\% & 49.21\% & 62.83\% \\
  \midrule
  \textbf{Dimension}
    & D1 & Documentation quality
    & 93.33\% & 95.30\% & 96.60\% & 95.95\% \\
    & D2 & Environment reconstruction
    & 83.33\% & 86.43\% & 91.67\% & 88.97\% \\
    & D3 & Workflow executability
    & 80.56\% & 86.46\% & 79.05\% & 82.59\% \\
    & D4 & Claim support
    & 80.56\% & 92.22\% & 74.77\% & 82.59\% \\
    & D5 & Reusability
    & 77.78\% & 78.26\% & 54.55\% & 64.29\% \\
  \cmidrule{2-7}
    & \multicolumn{2}{l}{\textit{Mean over D1--D5}}
    & \textit{83.11\%} & \textit{87.73\%} & \textit{79.33\%} & \textit{82.88\%} \\
  \bottomrule
  \end{tabularx}
\end{table*}

\paragraph{Datasets}
Prior work has reported that a portion of badged artifacts become inaccessible or non-functional over time~\cite{timperley2021understanding, muttakin2026state}. We therefore chose to re-execute the evaluation process rather than rely solely on previously awarded badges as ground truth. From the 642-paper pool in \autoref{sec:rubric}, we excluded calibration samples and artifacts with broken links, yielding 369 artifacts. To cover the judgment boundaries of AE while keeping manual review costs manageable, we performed a pre-labeling pass based on manual inspection without full evaluation. We stratified the pool into 104 cases that likely reach the Functional threshold, 246 boundary cases whose status cannot be determined by initial inspection, and 19 negative cases that clearly fail the Availability gate. Then we drew a stratified sample of 60 artifacts, 30 predicted to at least Functional, 20 boundary samples, and 10 predicted to fail Available. Two authors independently evaluated all 60 artifacts using the finalized \rubric{} and assigned dimension scores, gate decisions, and final badge recommendations. Disagreements were then adjudicated through discussion with a third author based on the recorded review evidence. The initial independent badge-level judgments achieved a Cohen's kappa of 0.85~\cite{5584447}, indicating strong inter-rater agreement. The final distribution consisted of 6 \textit{no-badge}, 20 \textit{Available-only}, 13 \textit{Available+Functional}, and 21 \textit{Available+Functional+Reusable}. Under the Functional binary threshold, 34 artifacts were positive, and 26 were negative.

\paragraph{Baselines}
We compare \tool{} against three representative coding agents and a project-execution agent. OpenHands~\cite{wang2025openhands} represents open-source general-purpose software engineering agents with strong repository navigation and development capabilities. Cline~\cite{cline2024} represents interactive coding assistants optimized for file-level editing and terminal use. Codex~\cite{openai2026codex} represents leading commercial coding agents with strong code understanding and generation capabilities. For Codex, we use a non-interactive configuration so that the full evaluation pipeline runs without human intervention. ExecutionAgent~\cite{executeagent2025} represents project-execution agents designed to make repositories runnable and testable. For all baselines, we keep the agent framework unchanged, inject the corresponding review protocol as task guidance, and enlarge step budgets to reduce premature termination.

\paragraph{Metrics}
We evaluate the system at three levels. At the badge level, we report exact-match accuracy and macro-averaged precision, recall, and F1 across the four badge classes. At the gate level, we treat Available, Functional, and Reusable as nested binary decisions and report precision, recall, and F1 for each. At the dimension level, we report four-class accuracy on the R/WR/WA/A scale. We also report a binarized version of the dimension-level results, collapsing R/WR into negative and WA/A into positive, and computing the corresponding precision, recall, and F1. Finally, we use the Functional threshold as an artifact-level binary separator, treating artifacts that meet it as positive and those that do not as negative, to assess whether the system can distinguish functionally valid artifacts from invalid ones.

\begin{table*}[t]
\centering
\caption{RQ2: Comparison with baselines. Values are means across three runs; bold indicates the best result.}
\label{tab:rq2_baseline_comparison}
\footnotesize
\setlength{\tabcolsep}{3pt}
\renewcommand{\arraystretch}{1.12}
\begin{tabularx}{\textwidth}{ll *{8}{>{\centering\arraybackslash}X}}
\toprule
\multirow{2}{*}{Agent} & \multirow{2}{*}{Protocol}
& \multicolumn{4}{c}{Badge-level} & \multicolumn{4}{c}{Functional-level} \\
\cmidrule(lr){3-6} \cmidrule(lr){7-10}
& & \makecell{Acc.} & \makecell{Prec.} & \makecell{Rec.} & \makecell{F1}
& \makecell{Acc.} & \makecell{Prec.} & \makecell{Rec.} & \makecell{F1} \\
\midrule
\multirow{3}{*}{OpenHands}
& ACM v1.1       & 48.89\% & 48.18\% & 45.04\% & 43.10\% & 82.78\% & 84.56\% &
85.29\% & 84.78\% \\
& \rubric{}         & 56.11\% & 53.88\% & 51.11\% & 49.62\% & 82.78\% & 83.24\% &
87.25\% & 85.15\% \\
& Skills+\rubric{}   & 59.44\% & 64.74\% & 59.32\% & 61.38\% & 81.67\% & 83.65\% &
85.29\% & 84.44\% \\
\addlinespace[2pt]

\multirow{3}{*}{Cline}
& ACM v1.1       & 34.44\% & 37.41\% & 37.61\% & 33.91\% & 70.00\% & 76.61\% &
67.65\% & 71.84\% \\
& \rubric{}         & 52.78\% & 54.92\% & 58.04\% & 52.47\% & 76.11\% & 80.49\% &
76.47\% & 78.35\% \\
& Skills+\rubric{}   & 59.44\% & 67.50\% & 61.45\% & 64.05\% & 80.56\% & 83.33\% &
83.33\% & 83.40\% \\
\addlinespace[2pt]

\multirow{3}{*}{Codex}
& ACM v1.1       & 39.44\% & 41.81\% & 33.89\% & 34.76\% & 73.89\% & 71.65\% &
89.22\% & 79.44\% \\
& \rubric{}         & 55.56\% & 66.65\% & 52.76\% & 57.69\% & 76.67\% & 78.85\% &
80.39\% & 79.63\% \\
& Skills+\rubric{}   & 67.78\% & 71.92\% & 66.72\% & 68.71\% & \textbf{86.11\%} &
85.32\% & \textbf{91.18\%} & \textbf{88.13\%} \\
\addlinespace[2pt]

\multirow{3}{*}{ExecutionAgent}
& ACM v1.1       & 31.11\% & 37.40\% & 30.83\% & 29.72\% & 56.11\% & 68.57\% &
41.18\% & 51.27\% \\
& \rubric{}         & 33.89\% & 37.93\% & 36.52\% & 32.21\% & 65.56\% & 83.62\% &
50.00\% & 62.14\% \\
& Skills+\rubric{}   & 32.78\% & 44.67\% & 41.71\% & 30.85\% & 58.33\% & 94.41\% &
28.43\% & 43.22\% \\
\midrule
\multicolumn{2}{l}{\tool{}}
& \textbf{70.56\%} & \textbf{74.05\%} & \textbf{72.05\%} & \textbf{70.64\%}
& 82.78\% & \textbf{98.55\%} & 70.59\% & 82.23\% \\
\bottomrule
\end{tabularx}
\end{table*}

\paragraph{Implementation Details}
Due to cost considerations, we selected DeepSeek-V4 Flash as the underlying language model for all systems. For Codex, which natively operates through OpenAI infrastructure, we accessed DeepSeek-V4 Flash via CC-Switch~\cite{ccswitch2025} to align the model backend as closely as possible across agents. For other baselines, we used their default configurations with enlarged step budgets to prevent premature termination. OpenHands was configured with 150 max iterations and 10 retries, Cline with 150 max requests and no timeout, and ExecutionAgent with 150 max steps and 1 retry. \tool{} was configured with 150 max steps. Experiments were conducted on a machine equipped with NVIDIA A100-80G GPUs running Ubuntu 22.04. All results reported for RQ1 and RQ2 are three-run means. RQ3 instead reports stability statistics aggregated over the three repeated runs.

\subsection{RQ1: Agreement with Human Adjudication}
\label{subsec:rq1-agent-gold}

\tool{} achieves 70.56\% exact agreement with human adjudication at the badge level. Agreement is strongest for judgments grounded in directly observable evidence. The Availability gate attains 97.53\% recall, and D1 records the highest four-class accuracy at 93.33\%. These results suggest that \tool{} can reliably identify whether an artifact is reviewable and whether its documentation provides a sufficient basis for evaluation.

\begin{table*}[t]
\centering
\caption{Stability and cost across three runs. For each baseline, we report its best-performing protocol in RQ2. Cost is the mean API cost per artifact run.}
\label{tab:rq3_stability}
\small
\setlength{\tabcolsep}{4pt}
\renewcommand{\arraystretch}{1.08}
\begin{tabular}{@{}llrrrrr@{}}
\toprule
System & Protocol & \makecell{Badge\\consistency} & \makecell{Badge\\divergence} & \makecell{Functional\\agreement} & \makecell{Failure\\rate} & Cost \\
\midrule
OpenHands & Skills+\rubric{} & 35/60 & 6/60 & 41/60 & 8.33\% & \$0.043 \\
Cline & Skills+\rubric{} & 21/60 & 3/60 & 43/60 & 16.67\% & \$0.068 \\
Codex & Skills+\rubric{} & \textbf{37/60} & 3/60 & \textbf{49/60} & 10.00\% & \$0.044 \\
ExecutionAgent & \rubric{} & 12/60 & 15/60 & 37/60 & 45.00\% & \$0.093 \\
\midrule
\tool{} & N/A & 32/60 & \textbf{1/60} & 45/60 & \textbf{0.00\%} & \$0.125 \\
\bottomrule
\end{tabular}
\end{table*}

Mismatch cases are concentrated more heavily on Functional and Reusable awards. In both cases, \tool{} shows higher precision than recall, indicating a conservative tendency in granting stronger positive decisions. This pattern is also reflected at the dimension level. D1 and D2 remain relatively stable, whereas D3, D4, and especially D5 are weaker, with D5 recording the lowest four-class accuracy. The gap between dimension-level and final badge-level agreement is expected because the final badge is a cumulative decision, an error on any gate-critical dimension can propagate to the overall badge outcome. Taken together, these results suggest that the main remaining difficulty lies not in basic repository inspection or setup recognition, but in turning execution evidence into stronger judgments about workflow validity, claim support, and especially reuse.

\subsection{RQ2: Comparison with Baseline Agents}
\label{subsec:rq2-baselines}

\autoref{tab:rq2_baseline_comparison} compares \tool{} with four baseline agents under three guidance conditions: ACM v1.1 badging policy~\cite{acmBadgingV11}, \rubric{}, and Skills+\rubric{}. Here, Skills+\rubric{} denotes the stage-specific phase skills from \tool{} combined with \rubric{}. This comparison is intended to test how progressively stronger review guidance affects agent behavior under our rubric-guided AE task formulation, rather than to reproduce the original venue badge process. At the badge level, \tool{} achieves 70.56\% exact-match accuracy, exceeding the strongest baseline, Codex under Skills+\rubric{}, at 67.78\%. At the Functional binary level, Codex under Skills+\rubric{} attains higher accuracy and recall, whereas \tool{} reaches a precision of 98.55\%, the highest among all systems. This result indicates that \tool{} is less likely to translate partial execution success into unsupported Functional awards.

The protocol effect is clearest at the badge level. For OpenHands, Cline, and Codex, performance improves consistently as the guidance moves from ACM v1.1 to \rubric{} and then to Skills+\rubric{}. The gains are much larger for final badge prediction than for basic Functional discrimination, suggesting that the main value of the rubric and staged skills is not merely to help agents execute artifacts, but to help them organize evidence into more calibrated badge judgments. Trajectory analysis explains why Codex is the strongest baseline. Its largest advantage comes from D3 recovery behavior rather than from uniformly better judgment. In several mismatch cases, \tool{} follows one documented entry point, encounters a technical failure, and then commits to a lower score, whereas Codex continues exploring alternative documented paths and recovers executable evidence, which accounts for a substantial fraction of the gap between the two systems. The traces also explain why \tool{} still achieves the best overall badge-level result. Its phase constraints and packet checking make positive decisions more conservative, especially at the Functional gate. As a result, \tool{} produces fewer unsupported upgrades from partial execution evidence, even though this conservatism also causes it to miss some human-awarded positive cases. ExecutionAgent shows an opposite failure mode, that despite strong execution-oriented behavior, it does not reliably shift to evidence-centered review under \rubric{} or Skills+\rubric{}, indicating that execution ability alone is insufficient for AE.

Overall, RQ2 supports two conclusions. First, official badge descriptions alone do not provide enough structure for stable automated review. Both \rubric{} and Skills+\rubric{} improve general-purpose agents substantially. Second, AE performance depends not only on whether an agent can run artifacts, but on whether it can turn execution into a bounded, evidence-driven judgment process. 

\subsection{RQ3: Stability Across Repeated Runs}
\label{subsec:rq3-stability}

We select the best-performing protocol configuration for each baseline from RQ2 as the basis for stability analysis. \autoref{tab:rq3_stability} reports stability across three independent runs under these selected configurations. Codex under Skills+\rubric{} achieves the highest exact three-run badge consistency at 37/60 and the highest Functional agreement at 49/60, indicating the strongest outcome convergence among runs that completed successfully. \tool{} is slightly lower on exact consistency at 32/60, but it is the only system with a zero failure rate, and it shows the fewest cases of extreme disagreement, with only 1 artifact receiving three different badge outcomes. Therefore, its advantage lies less in producing identical labels on every run than in avoiding runtime breakdowns and large result swings.

The sources of instability differ across baselines. OpenHands under Skills+\rubric{} produces more cases with three different badge outcomes across runs, suggesting that different runs often follow different review paths. Cline under Skills+\rubric{} shows lower exact consistency but limited extreme divergence, indicating more frequent movement between adjacent badge levels. Its failure traces repeatedly show context overflow, a pattern also observed in Codex under Skills+\rubric{}. Codex benefits from shorter and more direct review trajectories, which helps explain its stronger consistency, but its failures still reveal sensitivity to long accumulated context and often appear as hard interruptions once the context budget is exceeded. We therefore interpret these failures not as artifact-specific difficulty, but as a context-management limitation arising from the interaction between long-horizon agent trajectories and the underlying model context bound.  ExecutionAgent under \rubric{} presents a different failure mode. It has the lowest exact consistency at 12/60, the highest badge divergence at 15/60, and a failure rate of 45.00\%. Trace inspection shows that these failures are dominated by workspace initialization conflicts, artifact clone failures, broken container working directories, corrupted screen sessions, and missing parseable badge outputs rather than by artifact-specific difficulty. This suggests that an agent optimized for project execution is not naturally suited to the long-horizon, staged, and auditable evidence review process required by AE.

Overall, RQ3 yields two findings. First, exact three-run consistency favors agents with shorter trajectories and more direct decisions, which explains the strength of Codex on this metric. Second, for artifact evaluation, the more important notion of stability is not only whether repeated runs return identical badges, but whether they complete reliably and keep result variation within a limited range. In this sense, \tool{} provides a more suitable stability profile for AE assistance.

\subsection{RQ4: Reviewer Study}
\label{subsec:rq4-user-study}
\begin{figure}[t]
\centering
\includegraphics[width=\linewidth]{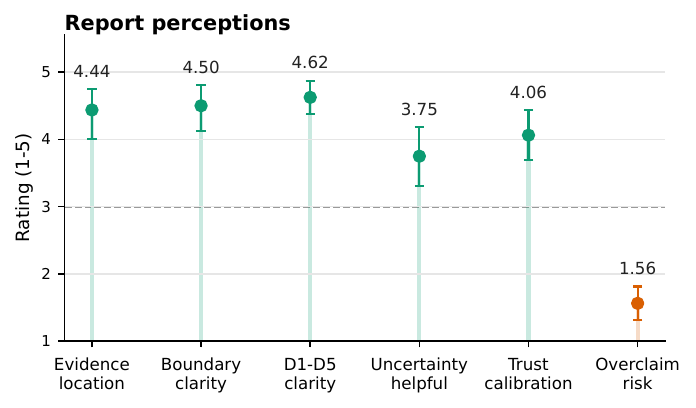}
\caption{Participant ratings of report qualities.}
\label{fig:rq4-report-perceptions}
\end{figure}
We conducted a controlled user study to examine how \tool{} reports support rubric-guided artifact evaluation. We selected 8 paper-artifact pairs from prior work in our lab and generated \tool{} reports for each and then prepared a with-report version and a no-report version. We recruited 8 participants with at least 3 years of software engineering research experience, none of whom had participated in the rubric design interviews. Following a Latin-square design, each participant completed 4 review tasks (2 with-report and 2 no-report), with each artifact evaluated under both conditions by different reviewers. All tasks used \rubric{}-guided AE, requiring participants to record review outcomes and complete scoring forms. The study yielded 32 valid review records in total.

Because of the limited sample size, we treat the results as descriptive comparisons. Self-reported confidence and all report perception items were collected on five-point Likert scales from 1 (lowest) as the lowest rating to 5 (highest) as the highest~\cite{nemoto2014likert}. Reviewers with access to the report reported higher confidence than those without it, with mean scores of 4.06 versus 3.63. \autoref{fig:rq4-report-perceptions} summarizes participants' assessments of the report itself. Evidence location received a mean score of 4.44 out of 5, boundary clarity 4.50, and dimension clarity 4.62. Uncertainty helpfulness and trust calibration were rated 3.75 and 4.06, while overclaim risk remained low at 1.56. These results suggest that the report was effective at helping reviewers locate evidence, understand what the current review did and did not establish, and follow the D1--D5 judgment structure. In the 16 with-report reviews, participant decisions agreed with the report in 15 cases. This suggests that the report was consistent with reviewer judgment in this controlled setting.

\section{Discussion}
\label{sec:discussion}
\subsection{Design Implications for Automated AE}
\label{subsec:discussion-automated-ae}
\begin{figure}[t]
\centering
\includegraphics[width=\linewidth]{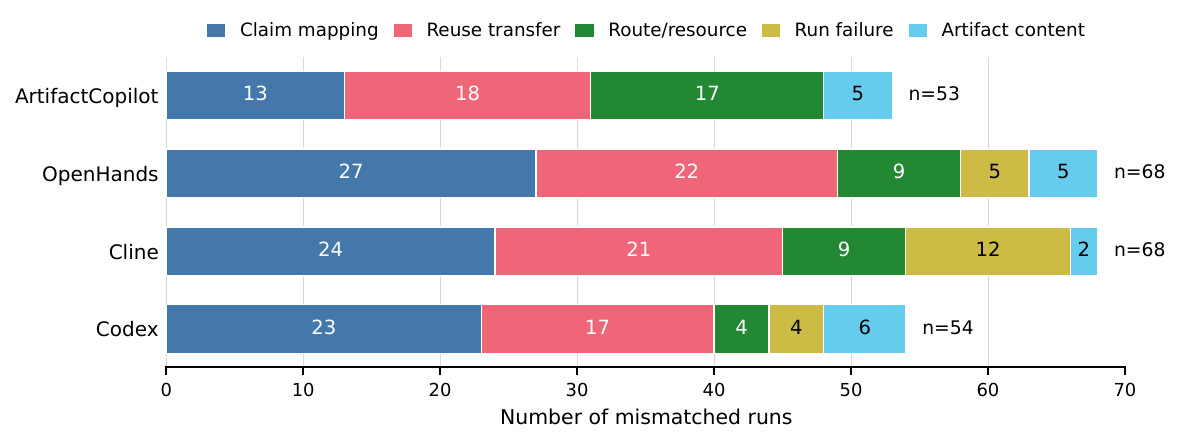}
\caption{Distribution of mismatch categories.}
\label{fig:discussion}
\end{figure}
Our mismatch analysis shows that the main barrier to automated artifact evaluation is no longer basic execution, but evidence interpretation. Across \tool{}, OpenHands, Cline, and Codex, many remaining errors arise after the artifact has already reached an executable state. We exclude ExecutionAgent from this analysis because most of its failures are infrastructure-level breakdowns before a usable AE judgment is produced, rather than interpretable review-decision mismatches. As \autoref{fig:discussion} shows, the dominant mismatch categories are claim mapping and reuse transfer, which together account for most disagreements. Both require evaluative judgments: whether an observed output is sufficiently connected to a paper claim, and whether a demonstrated use case constitutes genuine reuse rather than a minor workflow variation. This also exposes a limitation of our current system. Although \tool{} improves execution control and evidence organization, it still remains sensitive to execution context when interpreting evidence. The same artifact may lead to different badge outcomes depending on route choice, resource availability, or whether a repair is treated as acceptable intervention. These errors are therefore not merely implementation failures, but symptoms of the underlying subjectivity and context dependence of AE itself.

Taken together, these findings suggest that automated AE should not be framed as autonomous badge issuance from opaque execution outcomes. Instead, it should be designed as a reviewer-oriented evidence assistant that makes route choices explicit, records intervention burden, exposes uncertainty in claim mapping and reuse judgments, and localizes where human adjudication is still needed.

\subsection{Design Implications for Artifacts}
\label{subsec:discussion-artifact-design}
Our findings also suggest that \tool{} can serve as a detector of artifact quality, not only an evaluator. Artifacts that can be assessed stably by \tool{} are typically those with clearer structure, more explicit evidence, and better-defined review paths. Conversely, when the agent becomes uncertain or unstable, the cause often lies not in the code alone, but in underspecified artifact design. In this sense, evaluation difficulty is itself a useful signal of artifact quality.

This observation points to several concrete design practices for artifacts. First, artifacts should expose a canonical evaluation route, including the intended entry point, required resources, and acceptable bounded substitutes when full execution is too costly. Second, they should make outputs legible as evidence by explicitly linking scripts, logs, or tables to the corresponding paper claims. Third, artifacts targeting Reusable should provide at least one bounded transfer case with expected output, so that reuse can be validated rather than inferred. Finally, external dependencies such as GPUs, model downloads, API keys, or large datasets should be stated explicitly, together with a reduced validation path whenever possible. A high-quality artifact should therefore not be treated as merely a public code release, but should be designed and organized according to the standards of high-quality research software. Whether or not an artifact is submitted to AE, it should make its review path, the connection between outputs and paper claims, and its reusable validation path as explicit as possible, because these properties improve not only the stability of automated evaluation but also the practicality of human review, reproduction, and downstream use.

\subsection{Threats to Validity}
\label{subsec:discussion-validity}

The main internal threat lies in the construction of the adjudicated badges. Artifact evaluation inevitably involves judgment calls, especially when deciding whether an execution is representative, whether an output supports a paper claim, and whether a bounded transfer case establishes reuse. Because both the badges and \tool{} are grounded in the finalized \rubric{}, some degree of rubric-alignment bias cannot be fully excluded. We mitigate this threat through independent review, discussion, and third-party adjudication.

The main external threat lies in scope. Our 60-artifact suite was obtained by stratified random sampling to cover positive, negative, and difficult boundary cases, rather than to favor any system or estimate population frequency. The results also depend on the selected model backend, agent frameworks, and execution budgets. We therefore present \tool{} as a controlled instantiation, and view our findings as evidence about recurring challenges in AE automation rather than as universally generalizable performance results.

\section{Related Work}
\label{sec:related-work}

\subsection{Artifact Evaluation Practice and Standards}
\label{subsec:related-ae}

Prior work has examined different aspects of AE. Krishnamurthi et al.~\cite{krishnamurthi2013artifactEvaluation} discussed the early motivation and review process of AE at the conference level. Hermann et al.~\cite{hermann2020communityExpectations} analyzed reviewer expectations and showed that existing guidelines do not provide sufficiently uniform criteria for artifact judgment. Saucez et al.~\cite{saucez2019sigcommArtifacts} highlighted the tension between badge assignment and deeper validation. Timperley et al.~\cite{timperley2021understanding} studied recurring obstacles to reuse, including incomplete documentation and undeclared dependencies. These studies characterize AE practice, but they stop short of operationalizing badge policy into a usable scoring rubric. \rubric{} is designed to address this gap.

\subsection{LLM Agents and Automated Evaluation}
\label{subsec:related-agents}

LLM agents combine reasoning, tool use, and iterative feedback~\cite{yao2023react, Wang2023ASO, xi2025rise}. In software engineering, they have been applied to repository navigation~\cite{yang2024sweagent, zhang2024autocoderover}, general-purpose code development~\cite{wang2025openhands}, code repair~\cite{bouzenia2025repairagent}, automated project execution~\cite{executeagent2025}, vulnerability detection~\cite{Liu2025LargeLM, 309572}, and test generation~\cite{mundler2024code, AEGIS2024}. These systems are not targeting evidence-based reviews. In parallel, Pathak et al. showed that problem-specific rubrics significantly improve LLM code evaluation~\cite{prather2025rubric}, and the agent-as-a-judge paradigm extended agents to executable dynamic evaluation~\cite{zhuge2024agentAsJudge, gou2025mind2web2}. In AE, Heye et al. proposed an LLM-driven method for extracting reproducibility defects in cybersecurity artifacts~\cite{he2026llmAE}, but their work focuses on reproducibility signals rather than full badge-oriented review. More recently, Baek and Pradel proposed Artisan for agentic artifact evaluation~\cite{artisan2026}, focusing on validating paper-reported result objects such as tables against artifact outputs. Our work instead targets the complete AE workflow. \rubric{} and \tool{} support this workflow by combining a systematically constructed rubric with a phase-driven agent.

\section{Conclusion}
\label{sec:conclusion}
We introduced \tool{} and \rubric{}, arguing that AE automation is best understood as a rubric-guided evidence review task. \rubric{} translates descriptive badge standards into operational review criteria, while \tool{} realizes them through phase-specific review control. On 60 real artifacts with human-adjudicated badges, \rubric{} improves the AE performance of three coding agents over official policy prompts, and \tool{} achieves the highest exact badge agreement at 70.56\% while completing all repeated runs with zero failure rate. Our analysis shows that the main remaining challenge of automated AE is evidence interpretation rather than execution alone. It further suggests that higher-quality artifacts should make their review path, claim support, and reuse boundary explicit.

\bibliographystyle{IEEEtranS}
\bibliography{main}

@misc{acmBadgingV11,
  author       = {{Association for Computing Machinery}},
  title        = {Artifact Review and Badging -- Current},
  year         = {2020},
  month        = aug,
  note         = {Version 1.1, August 24, 2020. Accessed: June 17, 2026},
  howpublished = {\url{https://www.acm.org/publications/policies/artifact-review-and-badging-current}}
}

@article{krishnamurthi2013artifactEvaluation,
author = {Krishnamurthi, Shriram},
title = {Artifact evaluation for software conferences},
year = {2013},
issue_date = {April 2013},
publisher = {Association for Computing Machinery},
address = {New York, NY, USA},
volume = {48},
number = {4S},
issn = {0362-1340},
url = {https://doi.org/10.1145/2502508.2502518},
doi = {10.1145/2502508.2502518},
journal = {SIGPLAN Not.},
month = jul,
pages = {17–21},
numpages = {5}
}

@inproceedings{winter2022retrospectiveAE,
author = {Winter, Stefan and Timperley, Christopher S. and Hermann, Ben and Cito, J\"{u}rgen and Bell, Jonathan and Hilton, Michael and Beyer, Dirk},
title = {A retrospective study of one decade of artifact evaluations},
year = {2022},
isbn = {9781450394130},
publisher = {Association for Computing Machinery},
address = {New York, NY, USA},
url = {https://doi.org/10.1145/3540250.3549172},
doi = {10.1145/3540250.3549172},
booktitle = {Proceedings of the 30th ACM Joint European Software Engineering Conference and Symposium on the Foundations of Software Engineering},
pages = {145–156},
numpages = {12},
location = {Singapore, Singapore},
series = {ESEC/FSE 2022}
}

@article{timperley2021understanding,
  title={Understanding and improving artifact sharing in software engineering research},
  author={Timperley, Christopher S and Herckis, Lauren and Le Goues, Claire and Hilton, Michael},
  journal={Empirical Software Engineering},
  volume={26},
  number={4},
  pages={67},
  year={2021},
  publisher={Springer},
 url= {https://doi.org/10.1007/s10664-021-09973-5}
}

@article{muttakin2026state,
  title         = {The state of open science in software engineering research: A case study of icse artifacts},
  author        = {Muttakin, Al and Mondal, Saikat and Roy, Chanchal K},
  journal       = {2026 IEEE/ACM 48th International Conference on Software Engineering (ICSE)},
  year          = {2026},
  eprint        = {2601.02066},
  archivePrefix = {arXiv},
  primaryClass  = {cs.SE},
  url           = {https://arxiv.org/abs/2601.02066}
}

@inproceedings{eurosys2025lessonsLearned,
author = {D'Elia, Daniele Cono and Doudali, Thaleia Dimitra and Giuffrida, Cristiano and Matos, Miguel and Payer, Mathias and Pirelli, Solal and Portokalidis, Georgios and Schiavoni, Valerio and Signorello, Salvatore and Vahldiek-Oberwagner, Anjo},
title = {Lessons Learned from Five Years of Artifact Evaluations at EuroSys},
year = {2025},
isbn = {9798400719585},
publisher = {Association for Computing Machinery},
address = {New York, NY, USA},
url = {https://doi.org/10.1145/3736731.3746152},
doi = {10.1145/3736731.3746152},
booktitle = {Proceedings of the 3rd ACM Conference on Reproducibility and Replicability},
pages = {108–120},
numpages = {13},
location = {
},
series = {ACM REP '25}
}

@inproceedings{hermann2020communityExpectations,
author = {Hermann, Ben and Winter, Stefan and Siegmund, Janet},
title = {Community expectations for research artifacts and evaluation processes},
year = {2020},
isbn = {9781450370431},
publisher = {Association for Computing Machinery},
address = {New York, NY, USA},
url = {https://doi.org/10.1145/3368089.3409767},
doi = {10.1145/3368089.3409767},
booktitle = {Proceedings of the 28th ACM Joint Meeting on European Software Engineering Conference and Symposium on the Foundations of Software Engineering},
pages = {469–480},
numpages = {12},
location = {Virtual Event, USA},
series = {ESEC/FSE 2020}
}

@article{saucez2019sigcommArtifacts,
author = {Saucez, Damien and Iannone, Luigi and Bonaventure, Olivier},
title = {Evaluating the artifacts of SIGCOMM papers},
year = {2019},
issue_date = {April 2019},
publisher = {Association for Computing Machinery},
address = {New York, NY, USA},
volume = {49},
number = {2},
issn = {0146-4833},
url = {https://doi.org/10.1145/3336937.3336944},
doi = {10.1145/3336937.3336944},
journal = {SIGCOMM Comput. Commun. Rev.},
month = may,
pages = {44–47},
numpages = {4}
}

@article{treude2026rethinking,
  title={Rethinking Artifact Evaluation for Software Engineering in the Age of Generative AI},
  author={Treude, Christoph and Poskitt, Christopher M and Hoda, Rashina},
  journal={arXiv preprint arXiv:2604.16306},
  year={2026},
  url = {https://arxiv.org/abs/2604.16306}
}

@article{artisan2026,
  title         = {Artisan: Agentic artifact evaluation},
  author        = {Baek, Doehyun and Pradel, Michael},
  journal       = {arXiv preprint arXiv:2602.10046},
  year          = {2026},
  url           = {https://arxiv.org/abs/2602.10046},
}

@misc{sigmobile_artifact_guidelines,
  author       = {{ACM SIGMOBILE}},
  title        = {{ACM SIGMOBILE Research Papers Artifact Evaluation Guidelines}},
  year         = {n.d.},
  note         = {Prepared by the Research Artifact Evaluation Advisory Committee. Accessed: June 17, 2026},
  howpublished = {\url{https://www.sigmobile.org/grav/about/artifact-guidelines}}
}

@inproceedings{
zhuge2024agentAsJudge,
title={Agent-as-a-Judge: Evaluate Agents with Agents},
author={Mingchen Zhuge and Changsheng Zhao and Dylan R. Ashley and Wenyi Wang and Dmitrii Khizbullin and Yunyang Xiong and Zechun Liu and Ernie Chang and Raghuraman Krishnamoorthi and Yuandong Tian and Yangyang Shi and Vikas Chandra and J{\"u}rgen Schmidhuber},
booktitle={Forty-second International Conference on Machine Learning},
year={2025},
url={https://openreview.net/forum?id=Nn9POI9Ekt}
}

@inproceedings{
gou2025mind2web2,
title={Mind2Web 2: Evaluating Agentic Search with Agent-as-a-Judge},
author={Boyu Gou and Zanming Huang and Yuting Ning and Yu Gu and Michael Lin and Weijian Qi and Andrei Kopanev and Botao Yu and Bernal Jimenez Gutierrez and Yiheng Shu and Chan Hee Song and Jiaman Wu and Shijie Chen and Hanane Nour Moussa and TIANSHU ZHANG and Jian Xie and Yifei Li and Tianci Xue and Zeyi Liao and Kai Zhang and Boyuan Zheng and Zhaowei Cai and Viktor Rozgic and Morteza Ziyadi and Huan Sun and Yu Su},
booktitle={The Thirty-ninth Annual Conference on Neural Information Processing Systems Datasets and Benchmarks Track},
year={2026},
url={https://openreview.net/forum?id=AUaW6DS9si}
}

@inproceedings{asee_peer_41236,
author = {Klein-Gardner, Stacy and Abts, Leigh and Goldberg, Gail},
title = "The Engineering Design Process Portfolio Scoring Rubric (EDPPSR) – Initial Validity and Reliability (Fundamental)",
booktitle = "2022 ASEE Annual Conference \& Exposition",
year = "2022",
month = "August",
address = "Minneapolis, MN",
publisher = "ASEE Conferences",
note = {https://peer.asee.org/41236},
number = {10.18260/1-2--41236},
  url          = {https://peer.asee.org/41236}
}

@misc{rudolph2025don,
  title={Don't believe the hype. AI myths and the need for a critical approach in higher education},
  author={Rudolph, Jurgen and Ismail, Fadhil and Tan, Shannon and Seah, Pauline},
  journal={Journal of Applied Learning \& Teaching},
  volume={8},
  number={1},
  pages={6--27},
  year={2025},
  publisher={Kaplan Business School Australia Sydney, NSW}
}

@article{vijaykumar2018common,
  author    = {Vijaykumar, T. N.},
  title     = {A Common Standard to Fix Our Review Process},
  journal   = {SIGARCH Blog},
  year      = {2018},
  month     = {apr},
  url       = {https://www.sigarch.org/a-common-standard-to-fix-our-review-process-and-oh-i-was-wrong-about-one-thing/}
}

@inproceedings{zhang2024autocoderover,
  author    = {Zhang, Yuntong and Ruan, Haifeng and Fan, Zhiyu and Roychoudhury, Abhik},
  title     = {{AutoCodeRover}: Autonomous Program Improvement},
  booktitle = {Proceedings of the 33rd ACM SIGSOFT International Symposium on Software Testing and Analysis},
  series    = {ISSTA 2024},
  pages     = {1592--1604},
  year      = {2024},
  publisher = {ACM},
  address   = {New York, NY, USA},
  doi       = {10.1145/3650212.3680384},
  url       = {https://doi.org/10.1145/3650212.3680384},
  isbn      = {9798400706127},
  location  = {Vienna, Austria}
}

@inproceedings{yang2024sweagent,
author = {Yang, John and Jimenez, Carlos E. and Wettig, Alexander and Lieret, Kilian and Yao, Shunyu and Narasimhan, Karthik and Press, Ofir},
title = {SWE-agent: agent-computer interfaces enable automated software engineering},
year = {2024},
isbn = {9798331314385},
publisher = {Curran Associates Inc.},
address = {Red Hook, NY, USA},
booktitle = {Proceedings of the 38th International Conference on Neural Information Processing Systems},
articleno = {1601},
numpages = {125},
location = {Vancouver, BC, Canada},
series = {NIPS '24},
url       = {https://dl.acm.org/doi/10.5555/3737916.3739517}
}

@inproceedings{
yao2023react,
title={ReAct: Synergizing Reasoning and Acting in Language Models},
author={Shunyu Yao and Jeffrey Zhao and Dian Yu and Nan Du and Izhak Shafran and Karthik R Narasimhan and Yuan Cao},
booktitle={The Eleventh International Conference on Learning Representations },
year={2023},
url={https://openreview.net/forum?id=WE_vluYUL-X}
}

@article{Wang2023ASO,
  title={A survey on large language model based autonomous agents},
  author={Lei Wang and Chengbang Ma and Xueyang Feng and Zeyu Zhang and Hao-ran Yang and Jingsen Zhang and Zhi-Yang Chen and Jiakai Tang and Xu Chen and Yankai Lin and Wayne Xin Zhao and Zhewei Wei and Ji-rong Wen},
  journal={Frontiers of Computer Science},
  year={2023},
  volume={18},
  url={https://api.semanticscholar.org/CorpusID:261064713}
}

@article{xi2025rise,
  title={The rise and potential of large language model based agents: A survey},
  author={Xi, Zhiheng and Chen, Wenxiang and Guo, Xin and He, Wei and Ding, Yiwen and Hong, Boyang and Zhang, Ming and Wang, Junzhe and Jin, Senjie and Zhou, Enyu and others},
  journal={Science China Information Sciences},
  volume={68},
  number={2},
  pages={121101},
  year={2025},
  publisher={Springer},
  url = {https://doi.org/10.1007/s11432-024-4222-0}
}

@article{xu2025llm,
  title={{LLM-Based Agents for Tool Learning: A Survey}},
  author={Xu, Weikai and Huang, Chengrui and Gao, Shen and Shang, Shuo},
  journal={Data Science and Engineering},
  pages={1--31},
  year={2025},
  publisher={Springer},
  url       = {https://doi.org/10.1007/s41019-025-00296-9}
}

@article{Liu2025LargeLM,
  title={Large language model-based planning agent with generative memory strengthens performance in textualized world},
  author={Junyang Liu and Wenning Hao and Kai Cheng and Dawei Jin},
  journal={Eng. Appl. Artif. Intell.},
  year={2025},
  volume={148},
  pages={110319},
  url={https://api.semanticscholar.org/CorpusID:276859598}
}

@inproceedings{bouzenia2025repairagent,
author = {Bouzenia, Islem and Devanbu, Premkumar and Pradel, Michael},
title = {RepairAgent: An Autonomous, LLM-Based Agent for Program Repair},
year = {2025},
isbn = {9798331505691},
publisher = {IEEE Press},
url = {https://doi.org/10.1109/ICSE55347.2025.00157},
doi = {10.1109/ICSE55347.2025.00157},
booktitle = {Proceedings of the IEEE/ACM 47th International Conference on Software Engineering},
pages = {2188–2200},
numpages = {13},
location = {Ottawa, Ontario, Canada},
series = {ICSE '25}
}

@inproceedings{OGHarn2025,
author = {Sherman, Gabriel and Nagy, Stefan},
title = {No Harness, No Problem: Oracle-Guided Harnessing for Auto-Generating C API Fuzzing Harnesses},
year = {2025},
isbn = {9798331505691},
publisher = {IEEE Press},
url = {https://doi.org/10.1109/ICSE55347.2025.00239},
doi = {10.1109/ICSE55347.2025.00239},
booktitle = {Proceedings of the IEEE/ACM 47th International Conference on Software Engineering},
pages = {165–177},
numpages = {13},
location = {Ottawa, Ontario, Canada},
series = {ICSE '25}
}

@inproceedings {309572,
author = {Fengyu Liu and Yuan Zhang and Jiaqi Luo and Jiarun Dai and Tian Chen and Letian Yuan and Zhengmin Yu and Youkun Shi and Ke Li and Chengyuan Zhou and Hao Chen and Min Yang},
title = {Make Agent Defeat Agent: Automatic Detection of {Taint-Style} Vulnerabilities in {LLM-based} Agents},
booktitle = {34th USENIX Security Symposium (USENIX Security 25)},
year = {2025},
isbn = {978-1-939133-52-6},
address = {Seattle, WA},
pages = {3767--3786},
url = {https://www.usenix.org/conference/usenixsecurity25/presentation/liu-fengyu},
publisher = {USENIX Association},
month = aug
}

@inproceedings{
mundler2024code,
title={Code Agents are State of The Art Software Testers},
author={Niels M{\"u}ndler and Mark Niklas Mueller and Jingxuan He and Martin Vechev},
booktitle={ICML 2024 Workshop on Foundation Models in the Wild},
year={2024},
url={https://openreview.net/forum?id=LfyYSHGQMZ}
}

@inproceedings{AEGIS2024,
author = {Wang, Xinchen and Gao, Pengfei and Meng, Xiangxin and Peng, Chao and Hu, Ruida and Lin, Yun and Gao, Cuiyun},
title = {AEGIS: An Agent-based Framework for Bug Reproduction from Issue Descriptions},
year = {2025},
isbn = {9798400712760},
publisher = {Association for Computing Machinery},
address = {New York, NY, USA},
url = {https://doi.org/10.1145/3696630.3728557},
doi = {10.1145/3696630.3728557},
booktitle = {Proceedings of the 33rd ACM International Conference on the Foundations of Software Engineering},
pages = {331–342},
numpages = {12},
location = {Clarion Hotel Trondheim, Trondheim, Norway},
series = {FSE Companion '25}
}

@inproceedings{
wang2025openhands,
title={OpenHands: An Open Platform for {AI} Software Developers as Generalist Agents},
author={Xingyao Wang and Boxuan Li and Yufan Song and Frank F. Xu and Xiangru Tang and Mingchen Zhuge and Jiayi Pan and Yueqi Song and Bowen Li and Jaskirat Singh and Hoang H. Tran and Fuqiang Li and Ren Ma and Mingzhang Zheng and Bill Qian and Yanjun Shao and Niklas Muennighoff and Yizhe Zhang and Binyuan Hui and Junyang Lin and Robert Brennan and Hao Peng and Heng Ji and Graham Neubig},
booktitle={The Thirteenth International Conference on Learning Representations},
year={2025},
url={https://openreview.net/forum?id=OJd3ayDDoF}
}

@inproceedings{ICLR2024_7274ed90,
 author = {Ruan, Yangjun and Dong, Honghua and Wang, Andrew and Pitis, Silviu and Zhou, Yongchao and Ba, Jimmy and Dubois, Yann and Maddison, Chris and Hashimoto, Tatsunori},
 booktitle = {International Conference on Learning Representations},
 editor = {B. Kim and Y. Yue and S. Chaudhuri and K. Fragkiadaki and M. Khan and Y. Sun},
 pages = {27031--27098},
 title = {Identifying the Risks of LM Agents with an LM-Emulated Sandbox},
 url = {https://proceedings.iclr.cc/paper_files/paper/2024/file/7274ed909a312d4d869cc328ad1c5f04-Paper-Conference.pdf},
 volume = {2024},
 year = {2024}
}

@article{jonsson2007use,
title = {The use of scoring rubrics: Reliability, validity and educational consequences},
journal = {Educational Research Review},
volume = {2},
number = {2},
pages = {130-144},
year = {2007},
issn = {1747-938X},
doi = {https://doi.org/10.1016/j.edurev.2007.05.002},
url = {https://www.sciencedirect.com/science/article/pii/S1747938X07000188},
author = {Anders Jonsson and Gunilla Svingby}
}

@article{braunclarke2006thematic,
author = {Virginia Braun and Victoria Clarke},
title = {Using thematic analysis in psychology},
journal = {Qualitative Research in Psychology},
volume = {3},
number = {2},
pages = {77--101},
year = {2006},
publisher = {Routledge},
doi = {10.1191/1478088706qp063oa},
URL = {     
        https://doi.org/10.1191/1478088706qp063oa
},
eprint = {     
        https://doi.org/10.1191/1478088706qp063oa
}
}

@misc{cline2024,
  author       = {{Cline Bot}},
  title        = {Cline Bot: AI Coding Assistant},
  howpublished = {\url{https://cline.bot/}},
  year         = {2024},
  note         = {Accessed: 2026-06-23}
}

@article{executeagent2025,
author = {Bouzenia, Islem and Pradel, Michael},
title = {You Name It, I Run It: An LLM Agent to Execute Tests of Arbitrary Projects},
year = {2025},
issue_date = {July 2025},
publisher = {Association for Computing Machinery},
address = {New York, NY, USA},
volume = {2},
number = {ISSTA},
url = {https://doi.org/10.1145/3728922},
doi = {10.1145/3728922},
journal = {Proc. ACM Softw. Eng.},
month = jun,
articleno = {ISSTA047},
numpages = {23}
}

@misc{openai2026codex,
  author       = {{OpenAI}},
  title        = {Codex},
  howpublished = {\url{https://openai.com/codex/}},
  year         = {2026},
  note         = {Accessed: 2026-06-28}
}

@INPROCEEDINGS{he2026llmAE,
  author={Heye, David and Kindermann, Karl and Decker, Robin and Lohmöller, Johannes and Belova, Anastasiia and Geisler, Sandra and Wehrle, Klaus and Pennekamp, Jan},
  booktitle={2025 IEEE International Conference on Big Data (BigData)}, 
  title={Supporting Artifact Evaluation with LLMs: A Study with Published Security Research Papers}, 
  year={2025},
  volume={},
  number={},
  pages={5077-5085},
  doi={10.1109/BigData66926.2025.11401815},
  url={https://doi.org/10.1109/BigData66926.2025.11401815}
  }

@misc{icse2026ae,
  title        = {{ICSE} 2026 Artifact Evaluation Track},
  howpublished = {\url{https://conf.researchr.org/track/icse-2026/icse-2026-artifact-evaluation}},
  year         = {2026},
  note         = {48th IEEE/ACM International Conference on Software Engineering,
                  Rio de Janeiro, Brazil, April 12--18, 2026}
}

@misc{ase2026ae,
  title        = {{ASE} 2026 Artifact Evaluation Track},
  howpublished = {\url{https://conf.researchr.org/track/ase-2026/ase-2026-artifact-evaluation}},
  year         = {2026},
  note         = {41st IEEE/ACM International Conference on Automated Software Engineering, 2026}
}

@misc{fse2026ae,
  title        = {{FSE} 2026 Artifacts Track},
  howpublished = {\url{https://conf.researchr.org/track/fse-2026/fse-2026-artifacts}},
  year         = {2026},
  note         = {The ACM International Conference on the Foundations of Software Engineering, 2026}
}

@misc{issta2026ae,
  title        = {{ISSTA} 2026 Artifact Evaluation Track},
  howpublished = {\url{https://conf.researchr.org/track/issta-2026/issta-2026-artifact-evaluation}},
  year         = {2026},
  note         = {35th ACM SIGSOFT International Symposium on Software Testing and Analysis, 2026}
}

@inproceedings{prather2025rubric,
author = {Pathak, Aditya and Gandhi, Rachit and Uttam, Vaibhav and Ramamoorthy, Arnav and Ghosh, Pratyush and Jindal, Aaryan Raj and Verma, Shreyash and Mittal, Aditya and Ased, Aashna and Khatri, Chirag and Nakka, Yashwanth and Devansh and Challa, Jagat Sesh and Kumar, Dhruv},
title = {Rubric Is All You Need: Improving LLM-Based Code Evaluation With Question-Specific Rubrics},
year = {2025},
isbn = {9798400713408},
publisher = {Association for Computing Machinery},
address = {New York, NY, USA},
url = {https://doi.org/10.1145/3702652.3744220},
doi = {10.1145/3702652.3744220},
booktitle = {Proceedings of the 2025 ACM Conference on International Computing Education Research V.1},
pages = {181–195},
numpages = {15},
location = {
},
series = {ICER '25}
}

@inproceedings{nemoto2014likert,
  title        = {Developing Likert-Scale Questionnaires},
  author       = {Nemoto, Tomoko and Beglar, David},
  booktitle    = {JALT2013 Conference Proceedings},
  year         = {2014},
  editor       = {Sonda, N. and Krause, A.},
  pages        = {1--8},
  address      = {Tokyo},
  publisher    = {JALT},
  url          = {https://jalt-publications.org/sites/default/files/pdf-article/jalt2013_001.pdf}
}

@software{ccswitch2025,
  author = {Farion},
  title = {CC-Switch},
  url = {https://github.com/farion1231/cc-switch},
  year = {2025},
  note = {Accessed 2026-06-29}
}

@misc{anthropic2026agentskills,
  author       = {{Anthropic}},
  title        = {Agent Skills},
  howpublished = {Anthropic Platform Documentation},
  year         = {2026},
  url          = {https://platform.claude.com/docs/en/agents-and-tools/agent-skills/overview},
  note         = {Accessed: 2026-06-29}
}

@INPROCEEDINGS{5584447,
  author={Vieira, Susana M. and Kaymak, Uzay and Sousa, João M. C.},
  booktitle={International Conference on Fuzzy Systems}, 
  title={Cohen's kappa coefficient as a performance measure for feature selection}, 
  year={2010},
  volume={},
  number={},
  pages={1-8},
  doi={10.1109/FUZZY.2010.5584447},
  url= {https://doi.org/10.1109/FUZZY.2010.5584447}
  }

@misc{ArtifactCopilot2026,
  title = {ArtifactGuide},
  howpublished = {\url{https://figshare.com/s/4a21d79f93f92d11f2a1}},
  note = {Accessed: 2026-07-01}
}

\end{document}